# Computational Analysis using Multi-ligand Simultaneous Docking of Withaferin A and Garcinol Reveals Enhanced BCL-2 and AKT-1 Inhibition

Pronama Biswas
*Department of Biological Sciences*
*Dayananda Sagar University*
Bengaluru, India
0000-0003-2950-1530

Dishali Mathur
*Department of Biological Sciences*
*Dayananda Sagar University*
Bengaluru, India
0009-0006-3258-7717

Jinal Dinesh
*Department of Biological Sciences*
*Dayananda Sagar University*
Bengaluru, India
0009-0009-7148-0498

Hitesh Kumar Dinesh
*Department of Biological Sciences*
*Dayananda Sagar University*
Bengaluru, India
0009-0003-3225-9906

Belaguppa Manjunath Ashwin Desai
*Department of ECE*
*Dayananda Sagar University*
Bengaluru, India
0000-0002-4144-6266

*Abstract*— Developing an effective medicine to combat cancer and elusive stem cells is crucial in the current scenario. Withaferin A and Garcinol, important phytoconstituents of *Withania somnifera* (Ashwagandha) and *Garcinia indica* (Kokum) respectively, known for their therapeutic efficiency, have been used for several decades for treating various disorders, because of their anti-cancerous, anti-inflammatory and anti-invasive properties. This study investigates the potentials of withaferin A and garcinol in inhibiting BCL-2 and AKT-1, crucial proteins contributing in cancer cell persistence by evading apoptosis, increased cell proliferation, and inflammation. Molecular docking techniques, including single docking and MLSD, were used to understand the binding interaction of the ligands with BCL-2 and AKT-1. MLSD highlighted inter-ligand interactions among withaferin A and garcinol, against BCL-2, with a binding affinity of -11.88 ± 0.12 kcal/mol, surpassing the binding affinity of venetoclax (-9.73 ± 0.1 kcal/mol) a commercial inhibitor of BCL-2. For AKT-1, the binding affinity of withaferin A and garcinol (-13.74 ± 0.08 kcal/mol) surpassed the binding affinity of melatonin (-7.24 ± 0.06 kcal/mol), a commercial inhibitor of AKT-1. The MLSD results highlight the combined effects of garcinol and withaferin A, highlighting the importance of considering both the interactions of the bioactive compounds in the development of new medicines and strategies targeting cancer and elusive stem cells.

*Keywords*— *AKT-1, BCL-2, Withaferin A, Multi-ligand simultaneous docking, Synergistic effect*

## I. Introduction

Cancer is a chronic and prevalent malignancy characterized by various molecular alterations attributing to a state of uncontrolled cell division. As per WHO statistics (2020), cancer ranks as the second leading cause of death worldwide, with approximately 8.2 million fatalities annually. By 2030, this number is anticipated to rise to 11.5 million [1]. Therefore, understanding the molecular mechanisms involved in the progression of this disease is crucial for developing therapeutic drugs for effective treatment. Recent studies revealed the upregulation of PI3K/AKT/mTOR signaling pathway and apoptotic evasion as important mechanisms in various cancer types, resulting in metastasis, cellular proliferation, angiogenesis, and chemotherapy resistance[2].

Protein kinase B (PKB/AKT) belongs to ACG family of serine-threonine kinases regulating various cellular pathways and is often overexpressed in cancer. AKT-1, AKT-2, and AKT-3 are isoforms encoded by distinct genes, sharing over 80% structural similarity. Each isoform possesses a catalytic domain, surrounded by a highly conserved N-terminal pleckstrin homology (PH) domain and a C-terminal regulatory region. PH domain of AKT binds to phosphatidylinositol lipids (PIP2/PIP3), resulting in phosphorylation of Ser473 and Thr308 by protein kinase-1 (PDK1) and mechanistic target of rapamycin (mTOR) complex 2 (mTORC2), respectively, which activates AKT. Activated AKT phosphorylates several downstream targets such as GSK3β, promoting cell proliferation via Wnt pathway, and enhancing the expression of apoptosis inhibitors like BCL-2, highlighting its potential as a therapeutic target for cancer treatment [1,2].

B cell lymphoma-2 (BCL-2) protein family modulates programmed cell death through mitochondrial intrinsic pathway. BCL-2 family consists of both anti-apoptotic and pro-apoptotic proteins. The proapoptotic proteins (BAX, BAK, BIK, NOXA) promote apoptosis by initiating mitochondrial outer membrane permeabilization (MOMP), leading to caspase activation and the release of cytochrome c. Conversely, anti-apoptotic proteins (MCL-1, BCL-XL, BCL-2, BCL-W) inhibit apoptosis, aiding in cell survival and tumor progression [3]. Overexpression of BCL-2 prevents apoptosis by forming a heterodimer with BAX that binds to the membrane of mitochondria and, blocks the release of cytochrome c [4]. Several inhibitors including venetoclax, an FDA-approved drug, effectively target BCL-2 by directly binding to its BH3 domain, modulating apoptosis. However, in BCL-2, nonsynonymous mutations have reduced venetoclax efficacy, highlighting the need for discovering new inhibitors for BCL-2 [5].

Withaferin A, a C-28 steroid lactone extracted from *Withania somnifera* (Ashwagandha) is an important bioactive compound exhibiting anti-cancerous, anti-invasive, and anti-inflammatory properties. Withaferin A treatment at a concentration of 20 µl, significantly reduced cell viability of HeLa and MCF-7 cervical carcinoma cells by targeting Survivin, BCL-2, BCL-XL, and XIAP, resulting in enhanced apoptosis rate [6]. Similarly, garcinol, a polyisoprenylated benzophenone with a β-diketone moiety and phenolic moiety is extracted from *Garcinia indica* (commonly known as kokum) and possesses antioxidative, antimicrobial, anti-cancer, anti-inflammatory, anti-obesity, and anti-aging properties, making it a potential candidate for treatment against various ailments. Garcinol treatment significantly reduced the AKT phosphorylation in the human ovarian cancer cell line OVCAR-3 [7].

Molecular docking is a valuable approach for visualizing the inhibition of specific proteins by phytochemicals, revealing ligand-protein interactions at the atomic level. Widely used in computer-aided drug design (CADD), docking facilitates the development of novel protein inhibitors by identifying binding sites and optimal ligand conformations based on binding energy [8]. However, conventional docking methods, which analyze ligands individually, do not capture the multiple molecular interactions often present in biological systems, such as those involving cofactors, ions, and water molecules. These interactions can significantly influence ligand binding modes and docking scores. Multi-Ligand Simultaneous Docking (MLSD) addresses this limitation by concurrently docking multiple ligands, thereby mimicking natural molecular recognition processes. This method provides valuable insights into complex interactions, including additive, competitive, and combined effects of ligand pairs on protein inhibition [9]. In this study, we conducted both single-ligand docking and MLSD to evaluate the synergistic effects of withaferin A and garcinol on oncogenic target proteins. Our findings on MLSD highlight potential dual-drug therapeutic strategies for inhibiting BCL-2 and AKT-1.

## II. METHODS

### A. Protein selection

Fig.1 shows summary of the methodology for docking. The three-dimensional structure of AKT-1(PDB ID: 4GV1) and BCL-2(PDB ID: 4WJ9) were obtained from RCSB PDB. The protein quality was evaluated using "PROCHECK". In PROCHECK, both "Procheck" and "Verify 3D" analyses were performed. The assessment criteria included a Verify 3D score exceeding 80% and a Ramachandran plot score above 90%

### B. Ligand selection

Two plant-based ligands, garcinol and withaferin A were selected for targeting crucial cancer proteins. Both the phytochemicals were analyzed for PAINS using SwissADME. The three-dimensional structures of garcinol and withaferin A were derived from PubChem in SDF format, with Compound IDs, 5490884, 265237, respectively. Additionally, for validation purposes, the three-dimensional structures of the commercial inhibitors for AKT-1 and BCL-2 were also obtained from PubChem in SDF format, with CIDs 896 (melatonin) and 49846579 (venetoclax), respectively.

### C. Tools

Open Babel (http://openbabel.org) was employed to convert ligand SDF (Structure Data File) files into PDB (Protein Data Bank) format for docking studies. The ligand preparation was carried out using AutoDockTools-1.5.7, which involved addition of gasteiger charges. The resulting ligand PDB file was saved in PDBQT format as "ligand.pdbqt". The protein preparation was also carried out using AutoDockTools-1.5.7, which involved removal of co-crystallised ligands and water molecules, addition of hydrogen and gasteiger charges, resulting receptor PDB file was saved in PDBQT format as "receptor.pdbqt". Using UCSF Chimera the grid dimensions were calculated and saved as "config.txt".

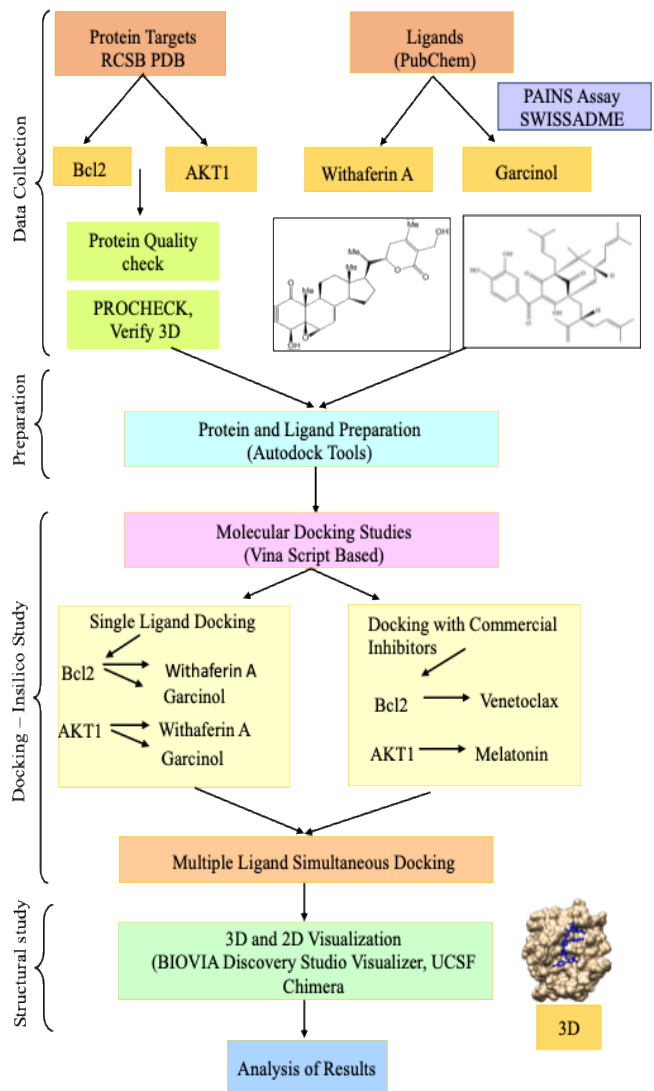

Fig. 1. Methodology followed for docking

Molecular docking was performed using the Vina Script method within a Conda environment accessed through Visual Studio Code. The resulting PDBQT files were analyzed in UCSF Chimera 1.17 and Discovery Studio 2021 Client to investigate binding pockets and amino acid interactions. The docking studies were executed on a server that featured 1.8 TiB storage capacity (1.7 TiB free), running on Ubuntu 22.04.1 LTS with a 6.5.0 kernel, 62 GiB of RAM, 40-core Intel Xeon Silver 4210 CPU (3.2 GHz max).

### D. Molecular Docking

In Single Docking, both the ligands and commercial inhibitors were individually docked with AKT-1, BCL-2 using Vina script method within Visual Studio Code. The resulting files were analyzed in UCSF Chimera to examine the 3D binding pockets, while two-dimensional visualizations were assessed in Discovery Studio 2021 Client to evaluate amino acid interactions between the ligands and proteins. In Multiple Ligand Simultaneous Docking (MLSD), the two ligands were docked concurrently with each protein (AKT1 and BCL-2). The resulting files were again analyzed in UCSF Chimera to explore the 3D binding pockets, while two-dimensional visualizations were evaluated in Discovery

Studio 2021 Client to evaluate both inter-ligand interactions along with amino acid interactions between the ligands and proteins.

## III. RESULTS

### A. Single Ligand Docking of BCL-2 and AKT-1

Withaferin A did not exhibit PAINS alerts, suggesting a low likelihood of nonspecific interactions with multiple targets. However, garcinol showed 1 PAINS ALERT (Catechol A). Withaferin A showed a better binding affinity with BCL-2 at -7.85 ± 0.06 kcal/mol than garcinol at -7.174 ± 0.033 kcal/mol (Table I). The 3D visualization of withaferin A revealed that it binds to a pocket similar to venetoclax, a commercial inhibitor for BCL-2. The 2D visualization revealed that withaferin A formed one hydrogen bond with ASN 143 (with a bond length of 4.699) along with hydrophobic interactions involving key amino acid residues (Gly 145, Ala 100, Tyr 108, Phe 104). On the other hand, the 3D visualization of garcinol revealed it also interacted with the same binding pockets forming hydrophobic interactions (Val133, Glu 136, Leu 137, Met 115) as venetoclax. The single ligand docking revealed that both ligands occupy the distinct parts of the venetoclax binding pocket, highlighting that the ligands could simultaneously bind, allowing for inter-ligand interactions, potentially resulting in increased stability within the pocket (Fig. 2A)

The single docking results of AKT-1 revealed that withaferin A and garcinol interacted with a stronger binding affinity of -9.68 ± 0.01 kcal/mol and -9.24 ± (-0.01) kcal/mol, respectively, as compared to commercial inhibitor, melatonin which had a value of -7.24 ± 0.06 kcal/mol (Table II). Garcinol formed a hydrogen bond with Lys179 (with a bond length of 4.25 Å) along with hydrophobic interactions with different key amino acid residues (Asp292, Val164) as melatonin, indicating that garcinol and melatonin share the same binding pocket. Similar results were observed for withaferin A as shown in Fig. 2(B). However, one unfavorable bond (with Lys158) was observed, and no hydrogen bond interactions were seen, revealing a weaker interaction of withaferin A with the melatonin binding pocket of AKT-1, compared to garcinol.

### B. Multi-Ligand Simultaneous Docking of BCl-2 and AKT-1

Single docking poses limitations when it comes to modeling a protein's flexibility during docking. In contrast, MLSD considers ligands based on their inter-ligand interactions, flexibility of protein binding pocket, and conformational orientations. The MLSD results showed that the binding affinity of the combination of ligands (-11.88 ± 0.12 kcal/mol) surpassed the binding affinity of venetoclax (9.73± 0.1 kcal/mol). both the ligands fit perfectly in the binding pocket of venetoclax, forming one pi-alkyl interaction between withaferin A and garcinol. Pi bonds are crucial as they help in stabilizing the conformation of ligands and intercalating the structure within the protein's binding pocket. The combined effects of ligands and contribution of pi-alkyl bonds and hydrophobic interaction with BCL-2 resulted in an enhanced binding affinity than venetoclax. Venetoclax interacts with the protein by forming hydrogen bonds with the amino acids TYR108, TRY202, ARG146, ASN143, TRP144. Furthermore, VAL133 forms pi-alkyl bonds, PHE104 forms pi-pi bonds and PHE112 forms a pi-sigma bond with venetoclax.

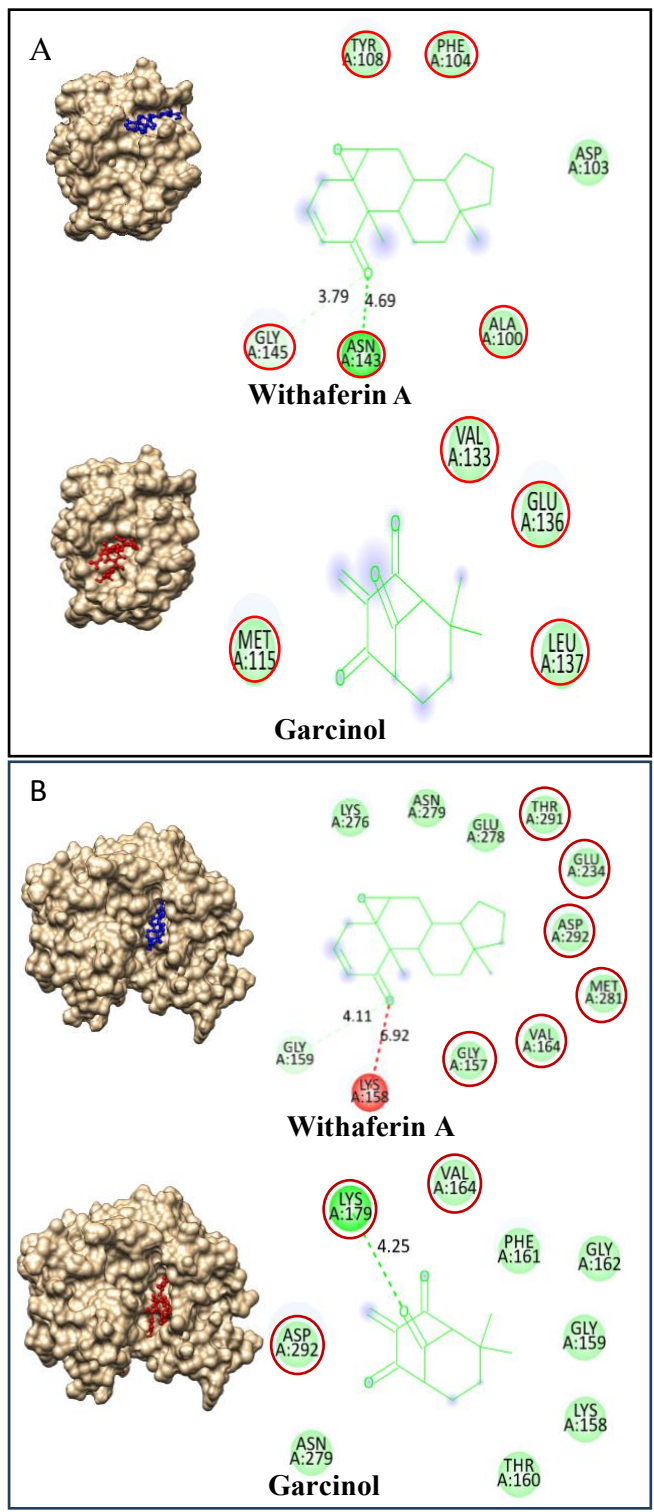

Fig. 2. Visualization of the binding pocket, amino acid interactions, and interligand interactions of single docking of BCL-2 and AKT-1. (A) Docking of a single ligand, withaferin A (blue), with Bcl-2 and garcinol (red), with BCL-2. (B) Docking of a single ligand, withaferin A (blue), with AKT-1 and garcinol (red), with AKT-1

The results of MLSD of BCL-2 are shown in Fig. 3. These key amino acids also interact with a combination of ligands forming pi-alkyl and hydrophobic interactions. These bonds prevent the binding of substrate and reduce the splitting of ligands with protein, which further can be studied using molecular dynamic simulations to confirm this stability. This further confirms that their binding pocket and target are similar to commercial inhibitors.

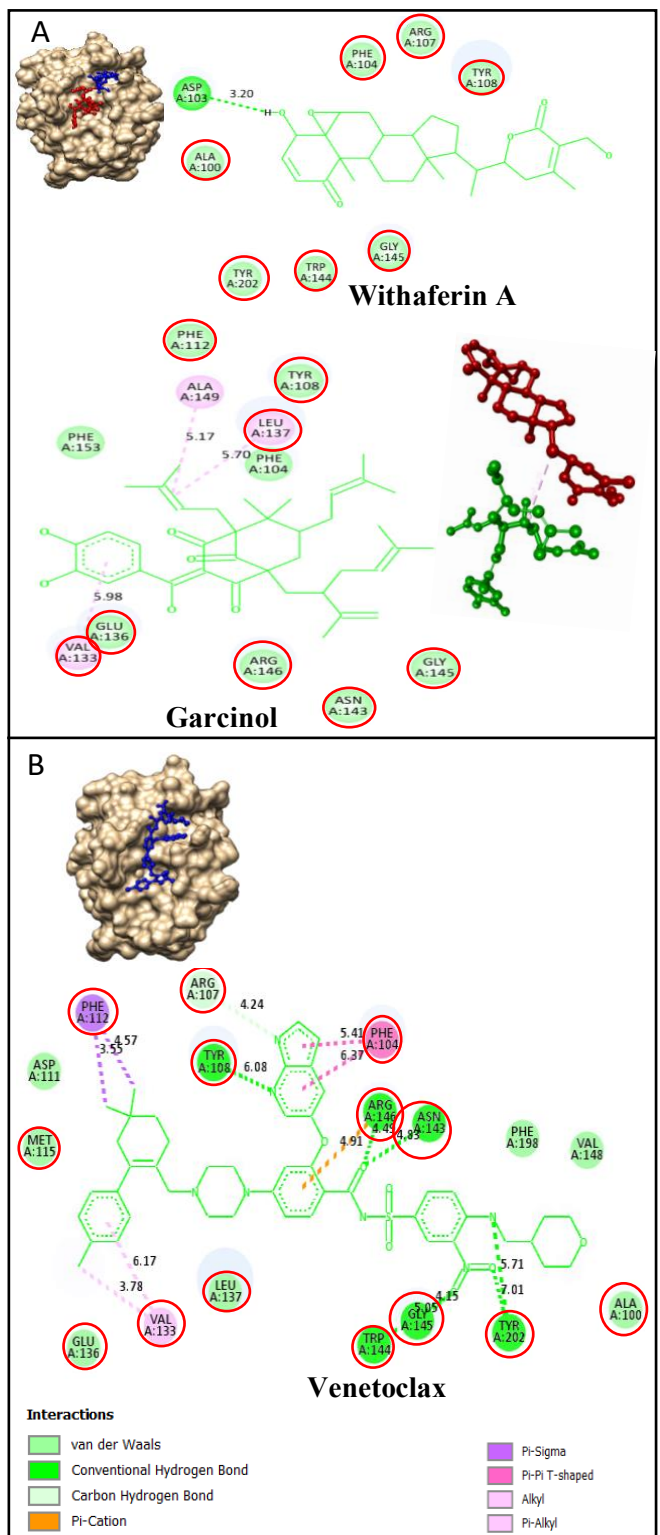

Fig. 3. Visualization of the binding pocket, amino acid interactions, and interligand interactions of MLSD of BCL-2 (A) Simultaneous docking of multiple ligands, withaferin A (blue) and garcinol (red), with BCL-2. (B) Docking of a single commercial inhibitor, venetoclax (blue), with BCL-2. The shared amino acids between the ligands and melatonin are highlighted in red circles, confirming their binding to the same pockets.

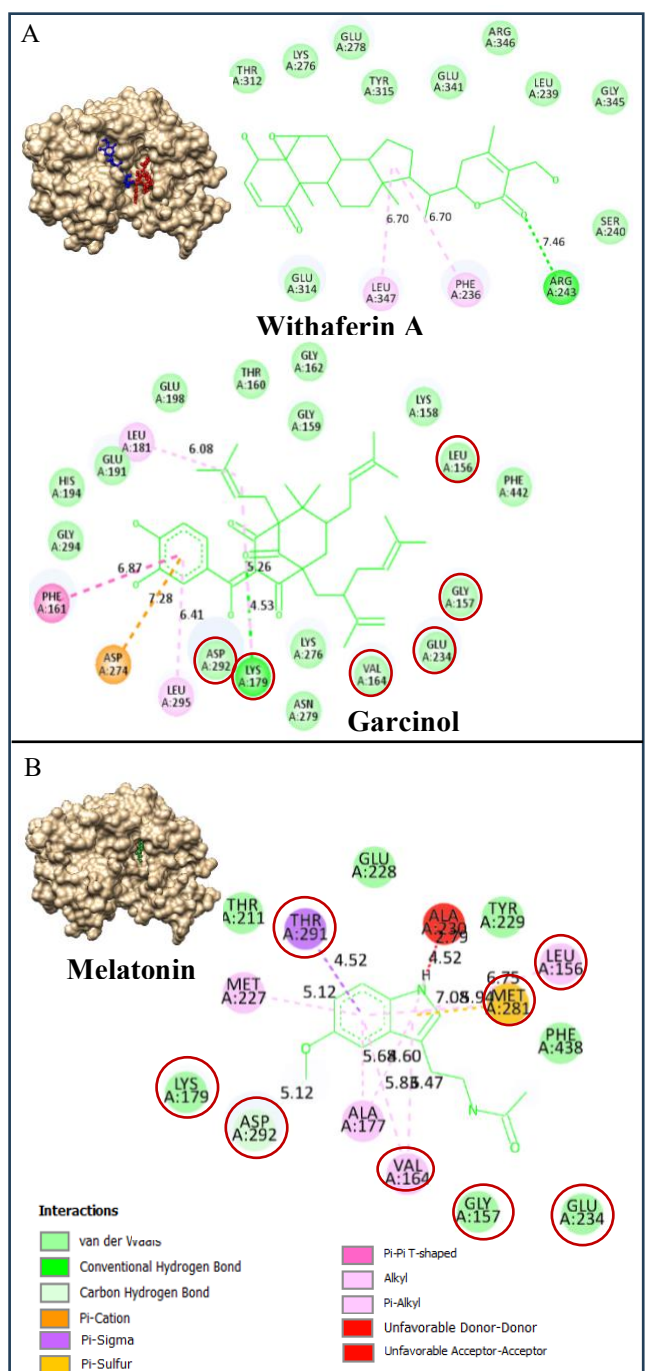

Fig. 4. Visualization of the binding pocket, amino acid interactions, and interligand interactions of MLSD of AKT-1 (A) Simultaneous docking of multiple ligands, withaferin A (blue) and garcinol (red), with AKT-1. (B) Docking of a single commercial inhibitor, melatonin (blue), with AKT-1. The shared amino acids between the ligands and melatonin are highlighted in red circles, confirming their binding to the same pockets.

MLSD results of AKT-1 exhibited significantly higher binding energy of the ligand combination (-13.74 ± 0.08 kcal/mol) compared to both single ligand docking and docking with melatonin. However, 3D pocket visualization indicated that garcinol binds to the melatonin binding pocket, while withaferin A was positioned adjacent to the melatonin binding pocket without any interactions with the amino acids of the original binding site as shown in Fig. 4. The amino acids interacting with melatonin also engaged with garcinol, forming hydrogen and pi-alkyl bonds (with Lys179) and hydrophobic interactions. These findings indicate that the enhanced binding affinity may be attributed to the combined effects of both compounds, potentially leading to an additive or synergistic interaction, as compared to melatonin.

TABLE I. BINDING AFFINITY OF Withaferin A, Garcinol AND MLSD

| Protein | PDB ID | Binding Affinity (kcal/mol) | | |
|---|---|---|---|---|
| | | Withaferin A | Garcinol | MLSD |
| BCL-2 | 8HOG | -7.85±0.06 | -7.17±0.03 | -11.88 ± 0.12 |
| AKT-1 | 4GV1 | -9.68±0.01 | -9.24 ± 0.013 | -13.74 ± 0.08 |

TABLE II. BINDING AFFINITY OF COMMERCIAL INHIBITORS

| Protein | Inhibitor | PubChem CID | Binding Affinity (kcal/mol) |
|---|---|---|---|
| BCL-2 | Venetoclax | 49846579 | -9.73 ± 0.1 |
| AKT-1 | Melatonin | 896 | -7.24 ± 0.06 |

## IV. Discussion

Cancer's global impact necessitates innovative solutions. Conventional chemotherapy and currently available therapeutics that target various cancers have limited therapeutic efficacy with a variety of side effects and drug resistance. Designing effective therapies is complicated by the heterogeneity of the tumor and its microenvironment. There is an urgent need to identify molecules with significant synergistic anti-cancerous activities, which could inhibit multiple proteins involved in the cancer cell pathway [10,11]. Synergistic effects occur when multiple bioactive agents interact to amplify their collective inhibitory impact, the resulting combined effect exceeds the impact produced by any single compound alone. This can arise by the formation of a stable intermolecular complex, dimers [10]. For example, a study by Alnuqaydan AM et al. showed that the synergistic anti-tumor effect of 5-fluorouracil and withaferin A in colorectal cancer cells is achieved through the induction of apoptosis and endoplasmic reticulum stress-mediated autophagy[11]. Our study revealed that both compounds possess the ability to inhibit both BCL-2 and AKT-1 highlighting their potential as effective anti-cancerous agents with synergistic therapeutic efficacy, additionally, a combination of these phytoconstituents has not been studied.

Withaferin A and garcinol displayed significant combined effects in inhibiting BCL-2 and AKT-1. The results of MLSD have justified this. The enhanced binding affinities observed in MLSD studies of BCL-2 are because of the inter-ligand interactions, including hydrogen bond formation. This interaction stabilizes the ligand complex within the BCL-2 binding pocket, preventing natural substrate binding. Moreover, both ligands successfully bind with the BH-3 domain of BCL-2 by forming hydrophobic interactions and pi-alkyl bonds with critical amino acids. This inter-ligand interaction was not observed in AKT-1. The multi-ligand simultaneous docking exhibited garcinol binding in the melatonin binding pocket. Hydrophobic interactions and hydrogen bonding contributed to the stabilization of garcinol binding. Hydrogen bonds are one of the strongest non-covalent interactions formed between the amino acids of protein and hydroxyl groups of ligands, further stabilizes the binding of ligands within the protein binding pocket. These findings suggest that both the ligands when combined, have a better potential to inhibit BCL-2 and AKT-1, important proteins involved in cancer cell pathways. The synergistic effects provide valuable insight into the use of these compounds for reshaping cancer treatment approaches and increasing efficacy by offering more profound effects on inhibiting cancer cell proteins.

The observed dual inhibition of proteins through the combination of ligands offers several advantages over synthetic drugs, including fewer side effects, and lower toxicity. The potential of these compounds to inhibit cancer cell proteins sheds light on the development of new cancer therapies, which could solve the drug resistance problems caused by synthetic drugs like venetoclax and melatonin. Strong evidence suggests that both the phytoconstituents provide health benefits, demonstrating significant anti-inflammatory, antioxidant, and anti-invasive properties.

This study has a few limitations. This study depends upon computational tools, which require experimental validations. The difference between the environment of computational docking and biological systems can profoundly impact the effectiveness and success of a drug. The structures in the databases are limited and there are chances they might not show the true confirmation because of the presence of missing residues. While performing docking studies, missing residues were not considered. Speculations made during docking studies can affect the results. Validating these results in biological systems will be essential for accurately assessing the therapeutic potential and any unforeseen interactions of these phytochemicals. Future studies should incorporate MD simulations to evaluate the stability and behavior of protein-ligand complexes, along with calculating the binding affinity using MM-PBSA. Additionally, toxicity assessment and ADMET profiling are crucial for evaluating the safety, effectiveness, and drug-like properties of potential therapeutic agents. Furthermore, experimental assays such as in-vitro studies on cancer cell lines should be performed to validate these findings and resolve any inconsistencies.

## CONCLUSION

Our study demonstrates that withaferin A and garcinol exhibit a synergistic inhibitory effect on key cancer-related proteins, BCL-2 and AKT-1, potentially enhancing apoptosis and reducing cancer cell proliferation. Their combined binding energies surpass those of venetoclax and melatonin, highlighting their therapeutic potential. These findings highlight a novel approach in cancer treatment, offering a potential strategy to overcome the limitations associated with conventional chemotherapy, including drug resistance and toxicity. Targeting multiple proteins using a combination of bioactive compounds could pave the way for developing effective and low-toxicity cancer therapies.

## Acknowledgment

The authors thank AIC-DSU for supporting them with the server for conducting their study. This study was supported by the Dayananda Sagar University Seed Grant. (DSU/REG/2022-23/740).